\documentclass[%
 reprint,
superscriptaddress,
 amsmath,amssymb,
 aps,
]{revtex4-1}

\usepackage{graphicx}% Include figure files
\usepackage{dcolumn}% Align table columns on decimal point
\usepackage{bm}% bold math
\usepackage{amsmath}
\usepackage{gensymb}
\usepackage{natbib}
\usepackage[english]{babel} 
\usepackage{color}
\usepackage{graphicx}% Include figure files
\usepackage{dcolumn}% Align table columns on decimal point
\usepackage{bm}% bold math
\usepackage{amsmath}
\usepackage{gensymb}
\usepackage{natbib}
\usepackage[english]{babel} 
\newcommand{\kms}{K$_2$MnS$_2$}
\newcommand{\kmse}{K$_2$MnSe$_2$}
\newcommand{\sse}{K$_2$MnS$_{2-x}$Se$_x$}

\newcommand{\Mn}{Mn$^{2+}$}
\newcommand{\K}{K$^{1+}$}

\newcommand{\Tn}{$T_\textrm{N}$}
\newcommand{\Tmax}{$T_{\textrm{max}}$}
\newcommand{\Jintra}{$J_{\textrm{intra}}$}
\newcommand{\Jinter}{$J_{\textrm{inter}}$}

\begin{document}

%\preprint{APS/123-QED}
\title{Incommensurate magnetism in \sse\ and prospects for tunable frustration in a triangular lattice of pseudo-1D spin chains }

\author{Ankita Bhutani}
\author{Piush Behera}
\author{Rebecca D. McAuliffe}
\affiliation{Department of Materials Science and Engineering, University of Illinois at Urbana-Champaign, Urbana, Illinois, USA.}
\author{Huibo Cao}
\author{Ashfia Huq}
\author{Melanie J. Kirkham}
\author{Clarina dela Cruz}
\affiliation{Neutron Scattering Division, Oak Ridge National Laboratory, Oak Ridge, TN 37831, United States}
\author{Toby Woods}
\affiliation{School of Chemical Sciences, University of Illinois at Urbana-Champaign, Urbana, Illinois, USA.}
\author{Daniel P. Shoemaker}
\affiliation{Department of Materials Science and Engineering, University of Illinois at Urbana-Champaign, Urbana, Illinois, USA.}
\date{\today}

\begin{abstract}
	
We report the first detailed investigation of \kms\ and \kmse\ from the \kms\ structure type and their magnetic solid solution \sse\ and find that compounds of this structure type consist of strongly-coupled pseudo-one-dimensional antiferromagnetic chains that collectively represent a frustrated two-dimensional triangular antiferromagnet. Bulk samples of K$_2$MnS$_{2-x}$Se$_x$ with $0 \leq x \leq 2$ are characterized using X-ray diffraction, neutron diffraction, magnetization and heat capacity measurements. An incommensurate cycloid magnetic structure with a magnetic propagation vector $k = [0.58~0~1]$ is observed for all samples in K$_2$MnS$_{2-x}$Se$_x$, and the ordering is robust despite a 12\% increase in cell volume. Geometric frustration of chains results in incommensurability along $a$ and a two-step magnetic transition. The varying geometries accessible in compounds of this structure type are presented as promising avenues to tune frustration.

\end{abstract}

\pacs{Valid PACS appear here}% PACS, the Physics and Astronomy
\maketitle

%\tableofcontents

\setlength{\parindent}{0.1 in}
\section{Introduction}

%https://pubs.acs.org/doi/pdf/10.1021/acs.inorgchem.8b00208
%1D superconductors
%\begin{enumerate}
%	\item{Bronger summary of these compounds}
%	\item{How he synthesized these compounds}
%	\item{no solid solution study done for this family}
%	\item{They have a TM, AFM nature, dimensionality in their structure}
%	\item{examples of 1D superconductors}
%	\item{mapping the phase diagram with x}
%	\item{mapping to understand change in magnetic ordering with doping}
%\end{enumerate}	 

The rich physics in the magnetism of low dimensional and frustrated systems has been a topic of great interest since last 4-5 decades. \cite{ramirez1994strongly, balents2010spin} Magnetic frustration, the competition of exchange couplings between localized spins, can be imposed by geometry or competing interactions and has broad implications for ground states and low-temperature properties. At low temperatures, peculiar behavior like spin ices \cite{bramwell2001spin} and spin liquids \cite{balents2010spin} can result when system fluctuates between different configurations or an ordered frustrated state with noncollinear and/or incommensurate magnetic structures can result with unsatisfied interactions in the Hamiltonian. \cite{nakamura2000incommensurate, kiryukhin2001magnetic}

Within such systems, frustration can arise with different
dimensionalities. Incommensurate magnetic ordering in quasi-1D magnetic systems can
be caused by frustrated \emph{intrachain} interactions as observed in 
LiCuVO$_4$, NaCu$_2$O$_2$, and CuCl$_2$, resulting in magnetic multiferroicity, \cite{gibson2004incommensurate,capogna2005helicoidal, banks2009magnetic, capogna2010magnetic, mourigal2011ferroelectricity} 
or it can be a hallmark of \emph{interchain} magnetic interactions resulting from frustrated arrangements of chains as observed in triangular-lattice antiferromagnets (TLAFs) such as CsNiCl$_3$ and Li$_2$NiW$_2$O$_8$. \cite{ranjith2016commensurate} 

\begin{figure}
	\centering\includegraphics[width=0.9\columnwidth]{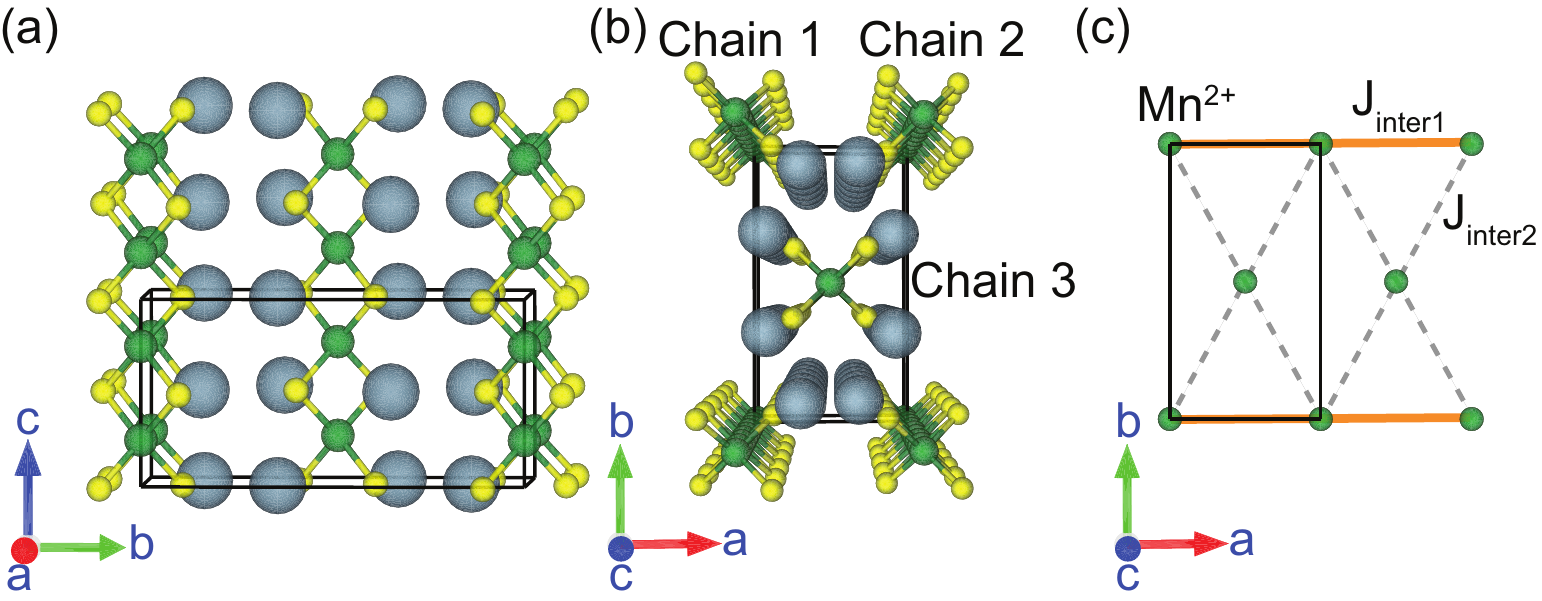}
	\caption{Crystal structure of 1D \kms\ and \kmse\ with views along $a$ and $c$ directions to
		show tetrahderally coordinated Mn-Ch chains running along $c$. \K\, \Mn\ and chalcogen
		are represented by grey, green and yellow spheres respectively.
	}
	\label{fig:unitcell} 
\end{figure}

In quasi-1D compounds, interchain coupling is required
to induce long-range three-dimensional ordering (a N\'{e}el temperature)
at finite temperatures. Exchange parameters $J_{ij}$ can represent the ``ideal'' nearest-neighbor ($nn$) and next nearest-neighbor ($nnn$) intrachain spin-exchange couplings, here
grouped as $J_\textrm{intra}$, while the longer-range interchain coupling between
the chains is represented by $J_\textrm{inter}$. (Figure \ref{fig:unitcell})
Materials that approach the ideal 1-D Heisenberg limit would be those where the
$J_\textrm{intra}$ is strong but $J_\textrm{inter}$ is weak; in this
case 3D magnetic ordering may be highly degenerate,
incommensurate, or not observed at all.

%An ideal one-dimensional magnetic system cannot be achieved in a three-dimensional 
%material with translational symmetry, because some degree of interchain coupling, however
%weak, will exist. This weak 
%coupling can induce long-range three-dimensional ordering (a N\'{e}el temperature)
%at finite temperatures in quasi-1D materials. 
%If the Hamiltoninan for such a quasi-1D magnetic system can be written as 
%$ H = -\sum J_{ij}\vec{S}_{i}\vec{S}_{j}$, 
%then the exchange parameters $J_{ij}$ can represent the ``ideal'' nearest-neighbor (nn) and next nearest-neighbor (nnn) intrachain spin-exchange couplings, here
%grouped as $J_{intra}$, while the longer-range interchain coupling between
%the three chains by $J_{inter1}$ and $J_{inter2}$ between chains 1 and 2, and 2 and 3 as shown in Figure \ref{fig:unitcell} (c).

%Incommensurate magnetic ordering in quasi-1D magnetic systems can
%be caused by frustrated intrachain interactions as observed in 
%LiCuVO$_4$, NaCu$_2$O$_2$, and CuCl$_2$ where superexchange is ferromagnetic, whereas the super-superexchange is antiferromagnetic and results in magnetic multiferroicity. \cite{gibson2004incommensurate,capogna2005helicoidal, banks2009magnetic, capogna2010magnetic, mourigal2011ferroelectricity} Research in low-dimensional magnetic systems containing Cu$^{2+}$ has also derived motivation from the need to understand cuprate superconductors. 
%Incommensurate magnetic ordering in quasi-1D magnetic systems can also be caused by competing interchain magnetic interactions resulting from spatial arrangement of magnetic chains as observed in triangular-lattice antiferromagnets (TLAFs) such as CsNiCl$_3$ and LiNiW$_2$O$_8$. \cite{ranjith2016commensurate} 
%It would therefore be interesting to study a system where the
%interactions are tunable and a range of spin and geometric constraints
%could be explored.

Identifying systems where exchange interactions are strong but 
frustrated is therefore of interest, with an additional benefit 
if the
interactions are tunable and a range of spin and geometric constraints
could be explored.
\kms\ and \kmse\ belong to a rich family of quasi-1D compounds
in the eponymous \kms\ structure type, where we concentrate
on the representatives where $A$$_2$$MX$$_2$, where $A$ = K, Na, Rb, Cs;
$M$ = Mn, Fe, Co, Zn, Si, Ge, Sn; and $X$ = S, Se, Te, P, As. 
\kms\ and \kmse\ were first synthesized by Bronger by the reaction of
potassium carbonate with manganese in a stream of hydrogen charged with 
chalcogen.\cite{bronger1989} Both these compounds exist in an orthorhombic space
group $Ibam$ with edge-sharing [Mn$X_4$]$^{-6}$ tetrahedral chains along 
the $c$ direction.
Viewed normal to the chains in Figure \ref{fig:unitcell}, the
chain locations form isosceles triangles with their short distance along
$\langle 100 \rangle$ and the long distance along $\langle 110 \rangle$.

%Mn-Mn 2.67 A in metallic manganese

The most closely-related compounds to the \kms\ structure type are
the family of $A$Fe$X_2$ ($A$ = K, Rb, Cs, Tl; $X$ = S, Se) compounds, which 
can be divided into three groups based on their crystal and magnetic structures. \cite{seidov2016magnetic} 
All $A$Fe$X_2$ compounds have tetrahedrally coordinated Fe$^{3+}$ chains along $c$, but many of these have an additional degree of structural freedom,
manifested as a monoclinic distortion (TlFeS$_2$, TlFeSe$_2$, KFeSe$_2$, and RbFeSe$_2$ with the magnetic moments ordered perpendicular to the chains; and KFeS$_2$, RbFeS$_2$ with the ordered moments
slightly tilted from the chain axis). \cite{asgerov2015magnetic, bronger1987antiferromagnetic} Only CsFeS$_2$ is orthorhombic, but it further distorts martensitically at its magnetic transition at 70~K, with moments slightly tilted off the chain direction.\cite{nishi1983observation}

%The most closely-related compounds to the \kms\ structure type are
%the family of $A$Fe$X_2$ ($A$ = K, Rb, Cs, Tl; $X$ = S, Se) compounds, which 
%can be divided into three groups based on their crystal and magnetic structures. \cite{seidov2016magnetic} All of these compounds have tetrahedrally coordinated Fe$^{3+}$ chains along $c$ direction, but many of these have an additional degree of structural freedom,
%manifested as a monoclinic distortion (TlFeS$_2$, TlFeSe$_2$, KFeSe$_2$, and RbFeSe$_2$ with the magnetic moments ordered perpendicular to the chains; and KFeS$_2$, RbFeS$_2$ with the ordered moments
%slightly tilted from the chain axis). \cite{asgerov2015magnetic, bronger1987antiferromagnetic} Only CsFeS$_2$ is orthorhombic which shows no magnetic order till 70K, at which point it undergoes a structural and magnetic transition with magnetic moments approximately along the chain direction. 

Unlike the well-studied $A$Fe$X_2$-type compounds, the compounds in the $A$$_2$$MX$$_2$ family
have not been examined in sufficient detail to understand any of their magnetic ground states.
Only the compound  K$_2$CoS$_2$ has been suggested to be a collinear antiferromagnet
(on the basis of neutron diffraction by Bronger, but without any published data). \cite{bronger1990ternare}
The antiferromagnetic nature of \kms\ and \kmse\ has been suggested based on susceptibility measurements. \cite{bronger1989}

%Till date, no magnetic structure or solid solution studies have been performed for $A_2$Mn$Ch_2$-type compounds. 

%\begin{figure*}
%	\centering\includegraphics[width=1.8\columnwidth]{fig-StructuralParameters}
%	\caption{\label{struc} Structural parameters obtained from laboratory powder X-ray diffraction and neutron powder diffraction (POWGEN) (a) Occupancy of Se (b) Volume of the unit cell (c) Intrachain distance between Mn1-Mn2 atoms (d) Interchain distance between Mn1-Mn3 atoms and Mn1-Mn4 atoms}
%\end{figure*} 

In this article, we present an investigation of the magnetic phase diagram of \sse\ using X-ray diffraction, neutron diffraction, magnetization and heat capacity measurements. This is the first detailed study of compounds from the family of $A$$_2$$MCh$$_2$-type compounds. We propose an incommensurate antiferromagnetic structure for \kms\ using single crystal neutron diffraction and powder neutron diffraction caused by geometric frustration of chains. All compounds in the \sse\ series show similar magnetic ordering and \Tn, and a two-step magnetic transition characteristic of TLAFs.

\section{Experimental Procedure} 

Bulk synthesis of the samples in the solid solution range of K$_2$MnS$_{2-x}$Se$_x$ with $0 \leq x \leq 2$ in increments of 0.2 was performed out using a solid state tube-in-a-tube method. Handling of reagents was performed in a glove box under argon.
% with O$_2$$<$0.6 and H$_2$O$<$0.6. 
Reactions were conducted by loading S and Se in 15 mm diameter quartz tubes in their nominal composition. Metallic K spheres and Mn powder were loaded in a smaller tube resting inside the bigger tube. These tubes were then sealed under vacuum using liquid nitrogen and reacted in box furnaces at 600$^\circ$C with a ramp rate of 1$^\circ$C/min and 48~h hold time, followed by furnace cooling. 
Synthesis of needle-shaped K$_2$MnS$_2$ single crystals was accomplished by high temperature gas flow, similar to the method reported by Bronger.\cite{bronger1989} 
K$_2$CO$_3$ and Mn powder were mixed using a mortar and pestle in 
a stoichiometric ratio and reacted under an incoming 5\% H$_2$/Ar 
stream charged with S vapor at 757$^\circ$C. 
%The high vapor pressure of hydrogen and sulfur mixture in the gas flow furnace facilitated the growth of single crystals of \kms\ in the shape of the needles.

%(a) Crystal structure of K$_2$MnS$_2$ (\textit{Ibam})perpendicular to a and c axis. The edge sharing Mn-Ch tetrahedra chain run along \textit{c} direction (K, gray; Mn, green; S, yellow). (b) Waterfall plot for laboratory X-ray diffraction patterns for \ks\, (x= [0,2] in increments of 0.2, except 0.2). Rietveld refinement difference curve of  is shown in the bottom in pink and indigo respectively. (c) Magnetic susceptibility for 

Powder X-ray diffraction (XRD) measurements were conducted in transmission with a Bruker D8 diffractometer with Mo-K$\alpha$ radiation. Rietveld analysis on X-ray diffraction patterns was carried out using TOPAS 5.~\cite{coelho2004topas} All samples were pure as viewed by X-ray analysis except for the $x=0.2$ sample. 
%X-ray fluorescence (XRF) data were collected using a Shimadzu EDX-7000 spectrometer under a He atmosphere. \dps{we do not show XRF?}

%Neutron diffraction measurements were carried out at 
%the Spallation Neutron Source (SNS) and High Flux Isotope Reactor 
%(HFIR) facilities at Oak Ridge National Laboratory (ORNL). 
%Time of flight data was collected on powder samples at beamline 11A on the POWGEN diffractometer at SNS for all compositions. The powder samples (1.5-2.5~g) were loaded in vanadium cans under He atmosphere. 
Neutron powder diffraction (NPD) was performed in vanadium cans on the POWGEN instrument at the
Spallation Neutron Source at Oak Ridge National Laboratory (ORNL).
The temperature was raised from 10K to a maximum of 50K with a ramp rate of 0.5K/min, and longer collections were taken at 10~K and 300~K. 
Processing and visualization of neutron powder diffraction data was done in the Mantid software.\cite{arnold2014mantid} 

Single crystal neutron diffraction was collected for \kms\ on the HB-3A four-circle diffractometer at the High Flux Isotope Reactor at ORNL, with a 
%To characterize the magnetic order in K$_2$MnS$_2$, we performed single crystal neutron diffraction at the HB-3A Four-circle diffractometer (FCD) equipped with a 2D detector at the High Flux Isotope Reactor(HFIR) at ORNL.
neutron wavelength of 1.550~\AA\ selected from a bent perfect Si-220 monochromator. \cite{chakoumakos2011four}
The selected crystal had a size of 0.9$\times$0.2$\times$0.5~mm and was sealed in a 0.7 mm diameter quartz tube (wall thickness 0.1 mm) to prevent air exposure. The crystal was held in place by another quartz capillary of 0.7 mm diameter.
%The sample is glued on the side of the quartz tube. 
The tube was mounted on the cold head of the closed cycle refrigerator (CCR)
to measure the temperature range from 4 K to 450 K.
Data were collected at 4.0~K and the (020) and (060) Bragg peaks were measured
at increasing temperatures. 
The nuclear and magnetic structure refinements and representation analysis are carried out with the FullProf Suite.\cite{rodriguez1990fullprof}
The magnetic symmetry analysis uses the Bilbao Crystallographic Server.\cite{aroyo_bilbao_2006,aroyo_bilbao_2006-1,aroyo_crystallography_2011,gallego_magnetic_2012,perez-mato_symmetry-based_2015}
% {if you like to include more details, you can mention how many nuclear reflections and magnetic reflections were collected …}

Magnetic susceptibility measurements were collected on a Quantum Design MPMS3 magnetometer. Heat capacity measurements were performed using a Quantum Design Dynacool Physical Property Measurement System (PPMS).
%, with chunks from reaction mixture mounted on the sample platform using N-grease. 
%Magnetic susceptibility and heat capacity measurements were carried out for $x=0, 1$ and $1.8$.

\begin{figure}
	\centering\includegraphics[width=0.95\columnwidth]{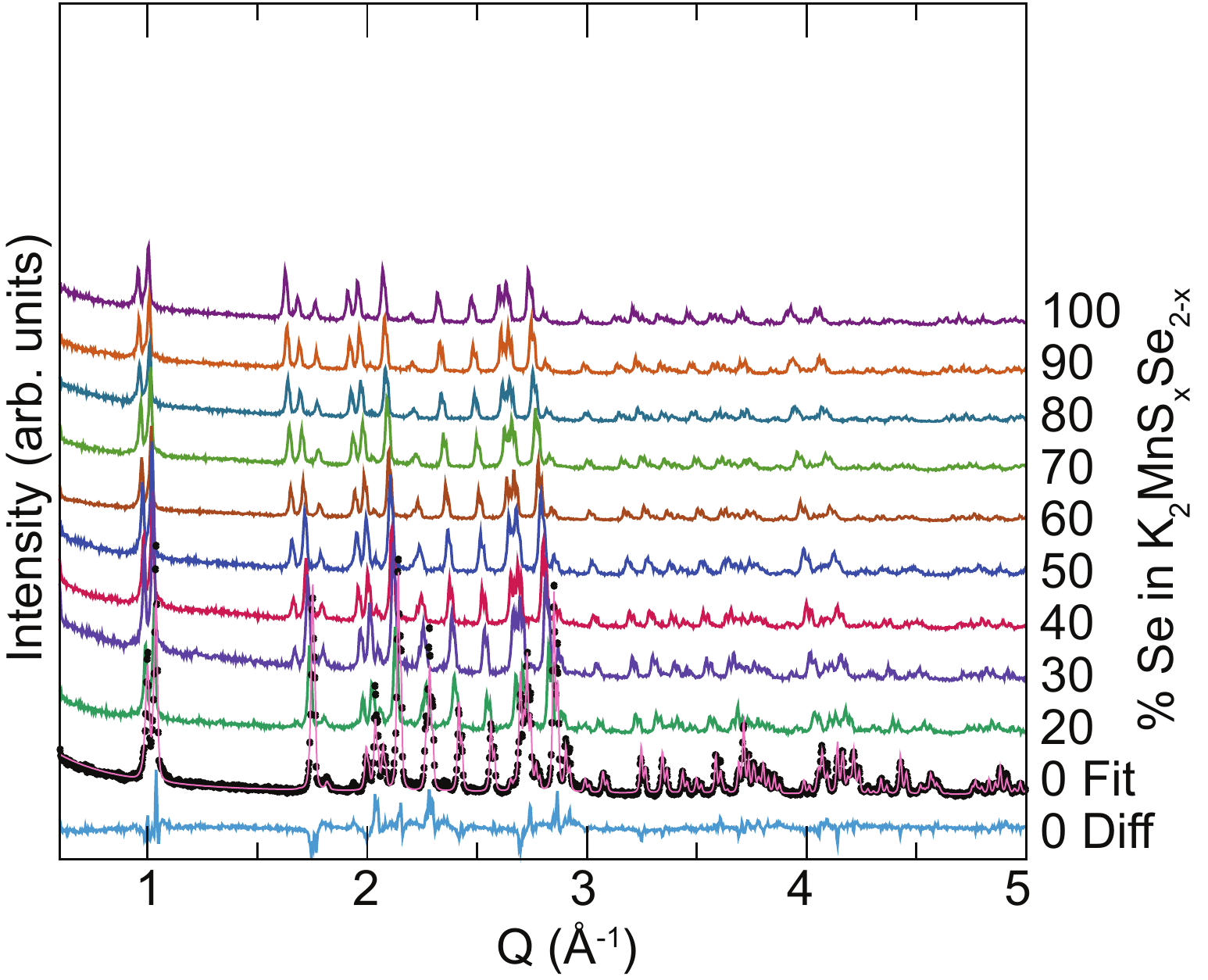}
	\caption{\label{xray} Laboratory X-ray diffraction patterns for K$_2$MnS$_{2-x}$Se$_x$ show consistent peaks for all samples, indicating a solid solution. The Rietveld refinement and difference curves for the K$_2$MnS$_2$ end member are shown at the bottom in pink and blue, respectively.}
\end{figure}

\begin{figure}[t]
	\centering\includegraphics[width=0.9\columnwidth]{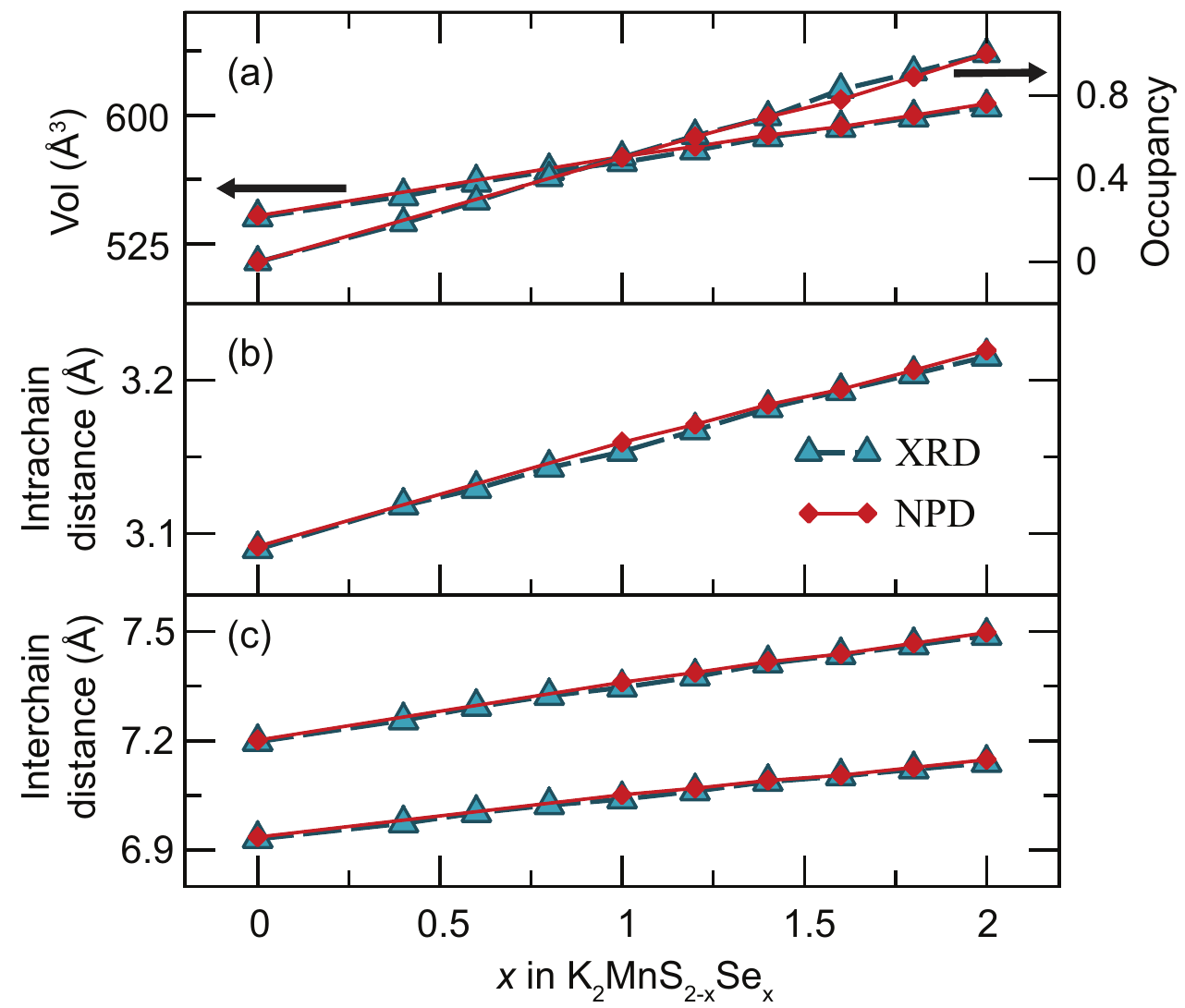}
	\caption{\label{fig:strucparam} Across the solid solution from \kms\ to \kmse, Rietveld refinements to XRD and NPD data show smoothly varying cell volume (a), Se occupancy (a), intrachain Mn--Mn distance (c), and short and long interchain distances (d).}
\end{figure}

\section{Results and Discussions}

%Quasi 1D antoferromagnets show a 1D short range ordering with a wide temperature dependence and a long range 3D ordering at a lower finite temperature. The magnetic intercations dominate along the chain direction (intrachain interactions) as compared to the interactions amonsgst the chains in the ab plane (inetrchain interactions)

%\subsection{SQUID}
%The magnetic suscpetibility data for K$_2$MnS$_x$Se$_{2-x}$ (x=[1,2] in increments of 0.2). ZFC and FC measurements 

\subsection{Confirmation of solid solution behavior from X-ray and neutron diffraction}
%Waterfall plot with the 0 or 100 percent pattern refinement
%\subsection{Structural variations in the solid solutions}
%\begin{enumerate}
%	\item{Bond length TM-Ch}
%	\item{Bond angles TM-Ch tetrahdedra}
%	\item{Volume}
%	\item{lattice parameters}
%	\item{Distance between TMs}
%	\item{Overlay all these graphs for both X-ray and neutron data with error bars to show the difference}
%	\item{}
%\end{enumerate}
%
%These compounds belong to the K2MnS2 structure type with an orthorhombic setting. The x-ray and neutron diffraction powder patterns collected are shown in Fig. The high vapor pressure of hydrogen and sulfur mixture in the gas flow furnace facilitated the growth of single crystals of K2MnS2 in the shape of the needles. X-ray diffraction measurment of these needles in the reflection geometry helped us in determining the orinetation of these needles. All peaks correspond to 110 reflection.  

Rietveld refinements to laboratory powder X-ray diffraction show all synthesized compositions K$_2$MnS$_{2-x}$Se$_x$ except the $x=0.2$ sample (Figure \ref{xray}). 
Figure \ref{fig:strucparam} shows the various refined structural parameters variation as a function of $x$ in K$_2$MnS$_{2-x}$Se$_x$ for both laboratory XRD and NPD. 
The Se occupancy varies linearly with $x$, for both XRD and NPD. 
Across the whole composition range, increasing the Se content toward \kmse\
leads to a 12\% volume increase.
The difference in ionic radii of S (1.84 \AA) and Se (1.98 \AA) \cite{shannon1969effective,shannon1970revised} and increasing covalent nature with $x$ causes an increase in Mn-S/Se bond length.
The increase in volume is mirrored by an increase in Mn-Mn distances--both intrachain and interchain as shown in Figure \ref{fig:strucparam} (b) and (c).

\begin{figure}[t]
	\centering\includegraphics[width=0.9\columnwidth]{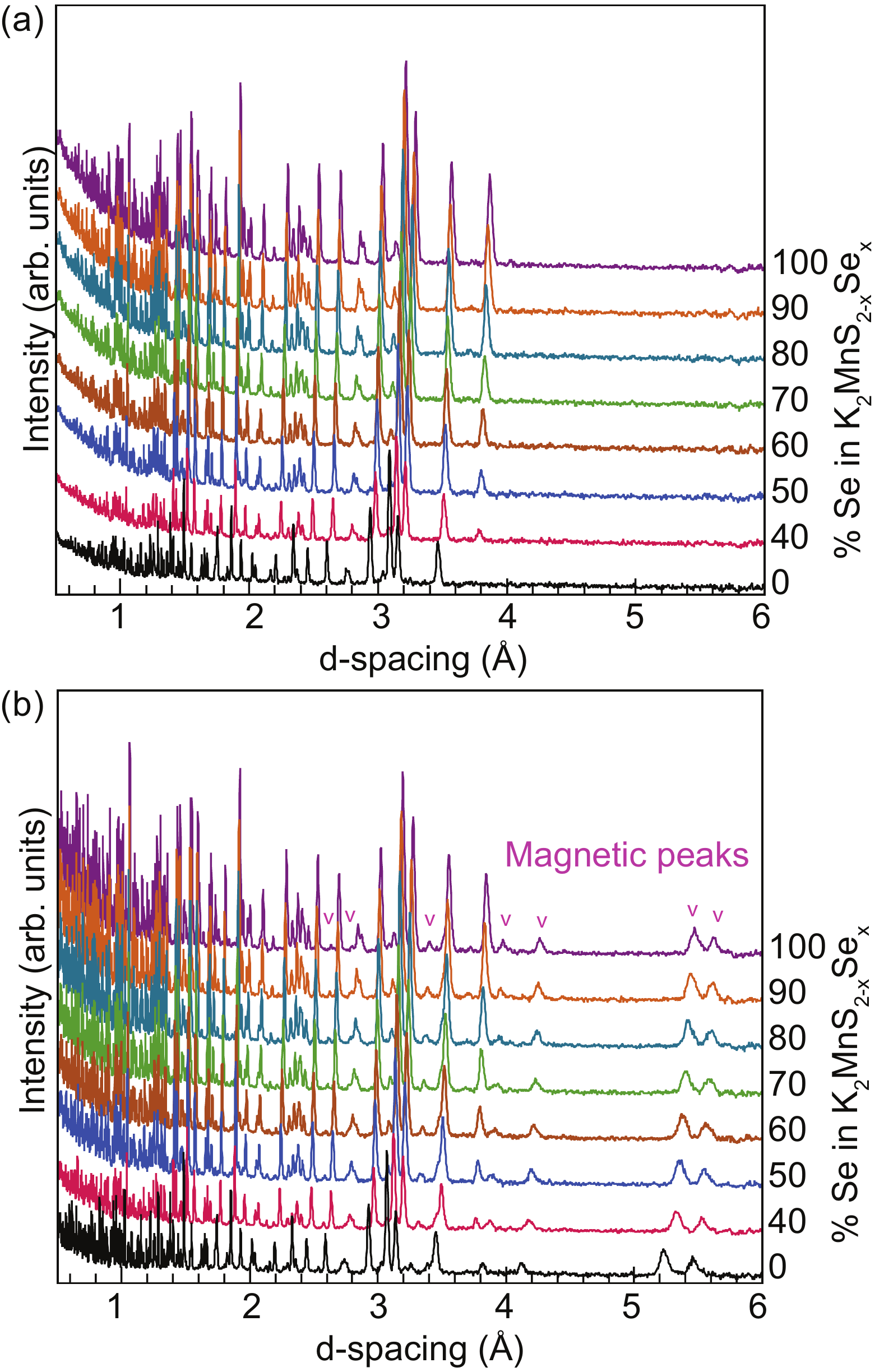}
	\caption{(a) Neutron powder diffraction patterns collected at (a) 300~K and
		(b) 4~K  for K$_2$MnS$_{2-x}$Se$_x$.
		The 4~K data show magnetic peaks for all compositions marked with the pink 
		pointers. The positions and intensities of the peaks portend equivalent
		magnetic ordering among the samples, but the propagation vector is
		incommensurate with the nuclear structure.
	}
	\label{fig:npd} 
\end{figure}

Figure \ref{fig:npd}(a,b) show the NPD patterns collected at
300~K and 10~K across the composition range of K$_2$MnS$_{2-x}$Se$_x$. 
The NPD data at 10~K show magnetic ordering in all compositions, with 
magnetic peaks that appear consistent across the compositional range,
despite the large volume change and change in inter/intrachain distances.
Surprisingly, the magnetic peaks could not be indexed to a commensurate
$k$-vector, including the [$0 0 \frac{1}{2}$] ordering proposed for 
K$_2$Co$S_2$ by Bronger.\cite{bronger1989}

\begin{figure*}
	\centering\includegraphics[width=2\columnwidth]{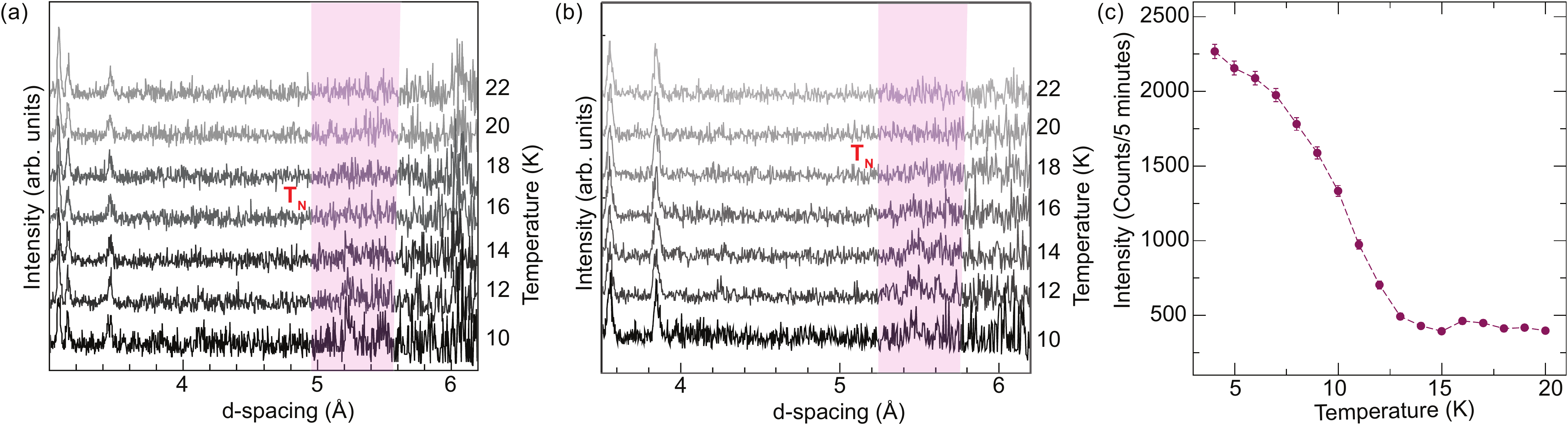} 
	\caption{NPD with increasing T shows disappearance of the magnetic peaks around 15 K for (a) \kms\ and (b) \kmse, while triple-axis single-crystal intensity on the $(0.42 1 1)$ peak (c) provides a more precise view of the order parameter evolution. The exact temperature is best judged by heat capacity measurements. }
	\label{fig:op} 
\end{figure*}

Initial magnetic structure predictions using SARAh and FullProf suggested incommensurate ordering of the Mn$^{2+}$ moments. Rough determination of $T_N$ from NPD can be accomplished by 
noting the temperature of disappearance of magnetic peaks at $d=5.2$~\AA. Figure \ref{fig:op} (a) and (b) show this range of NPD data for K$_2$MnS$_2$ and K$_2$MnSe$_2$ at 16$\pm$2K and 18$\pm$2K respectively. T$_N$ for all compositions lie within in this temperature range and do not show a dramatic increase with $x$. This is despite the increasing distance between intrachain Mn1-Mn2 ions (4$\%$), interchain Mn1-Mn3 ions (3$\%$) and interchain Mn1-Mn4 ions (4$\%$).

%\subsection{SEM micrographs}
%The high vapor pressure of hydrogen and sulfur mixture in the gas flow furnace facilitated the growth of single crystals of K$_2$MnS$_2$ in the shape of the needles. The SEM micrographs of K$_2$MnS$_2$ are presented in Figure. 

\subsection{Incommensurate magnetic ordering in \sse}

%\begin{figure}
%	\centering\includegraphics[width=0.6\columnwidth]{fig-Jpaths}
%	\caption{\label{fig:maginteractions} Magnetic interactions }
%\end{figure} 

\begin{figure}[b]
	\centering\includegraphics[width=0.8\columnwidth]{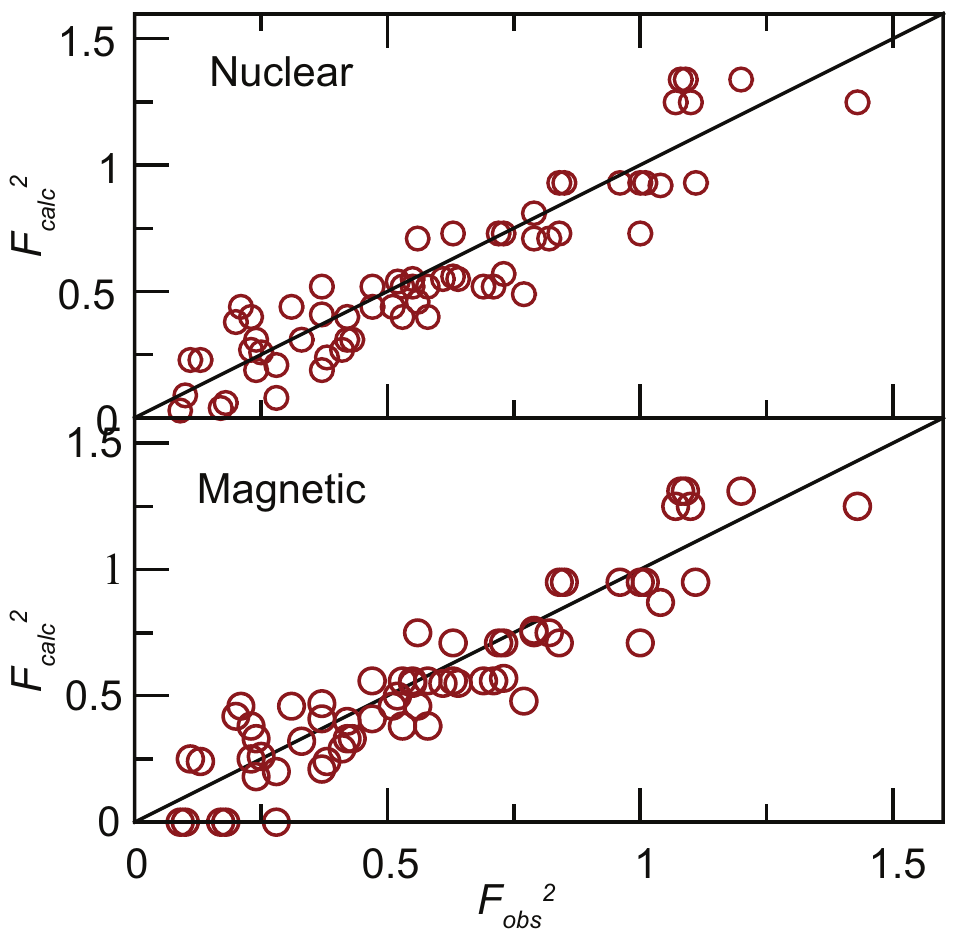}
	\caption{\label{fig:fobs} Calculated vs observed structure factors for nuclear and magnetic reflections obtained from single crystal neutron diffraction refinement of \kms}
\end{figure} 

\begin{figure}[b]
	\centering\includegraphics[width=0.8\columnwidth]{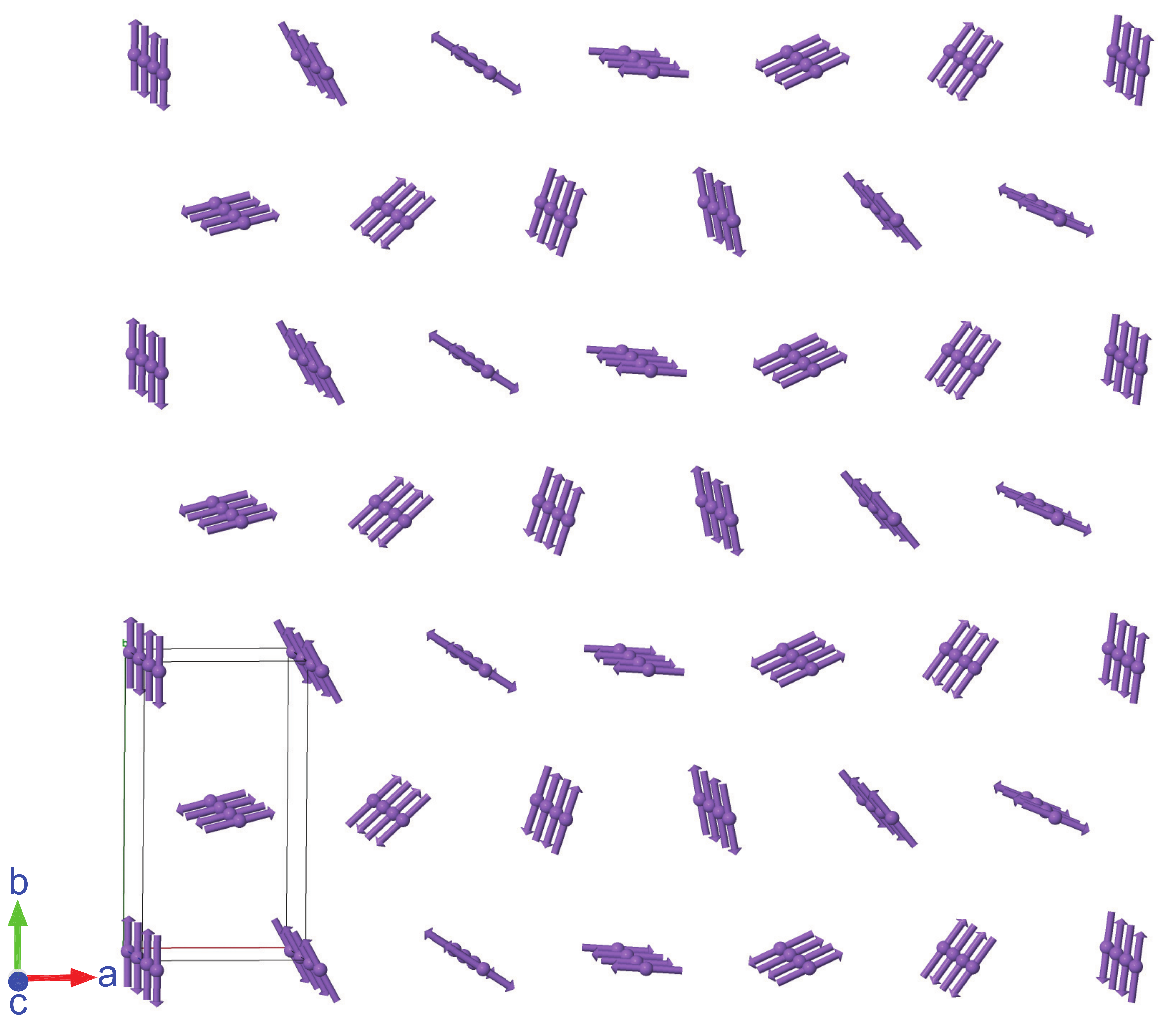}
	\caption{\label{fig:magstruc} Magnetic structure of \kms\ determined from single crystal neutron diffraction showing the cycloid arrangement of spins in the $ab$ plane with incommensurability along $a$ direction.}
\end{figure}

%The single crystal magnetic structure of 
%The main reflections of the neutron The single crystal 

Magnetic reflections appear below $T = 17$~K for all compositions in \sse\ as shown in Figure \ref{fig:op}(a,b). An order parameter plot for is shown in Figure \ref{fig:op}(c) from single crystal neutron diffraction of \kms. The presence of magnetic reflections in both single crystal and powder neutron diffraction is consistent with our magnetic susceptibility and heat capacity measurements.
The magnetic phase was indexed using an incommensurate magnetic propagation vector $k = [0.58~0~1]$ at 4~K.
The final refinement of single crystal neutron diffraction data is shown in Figure \ref{fig:fobs} where the observed and calculated structure factors for the nuclear and magnetic refinements are plotted. 
The refined magnetic structure results in a cycloidal configuration of magnetic moments in the $ab$ plane as shown in Figure \ref{fig:magstruc}. 
The spins are antiferromagnetically coupled along the chains in the $c$ direction. 
The total refined magnetic moment is 2.279~$\mu_\textrm{B}$, considerably smaller than 5.92~$\mu_\textrm{B}$ expected for \Mn\ ($S = 5/2)$.
A reduced magnetic moment is often observed in quasi-1D spin systems such as NaFeGe$_2$O$_6$, Na$_2$TiCl$_4$ and has been believed to be caused by spin fluctuations and covalency.\cite{ding2018unraveling,winkelmann1995magnetic} Delocalization of $d$ electrons could be another explanation for reduced magnetic moment as observed in RbFeSe$_2$. \cite{seidov2016magnetic} Here, it is more likely that the reduced moment arises from latent disorder due to the frustrated moments in \kms. 
Accordingly, the steady increase of the magnetic peak $(0.42~1~1)$ order parameter upon cooling to 4~K indicates that the magnetic ordering is not complete. (Figure \ref{fig:op} (c))
Magnetic refinements to the NPD data at 10~K confirm that the cycloid model can be assumed as the magnetic structure for the full substitution range (Figure \ref{fig:npd}). However, a very good fit for the powder intensity at $d = 5.23$~\AA\ could not be obtained.

\begin{figure}[t]
	\centering\includegraphics[width=0.8\columnwidth]{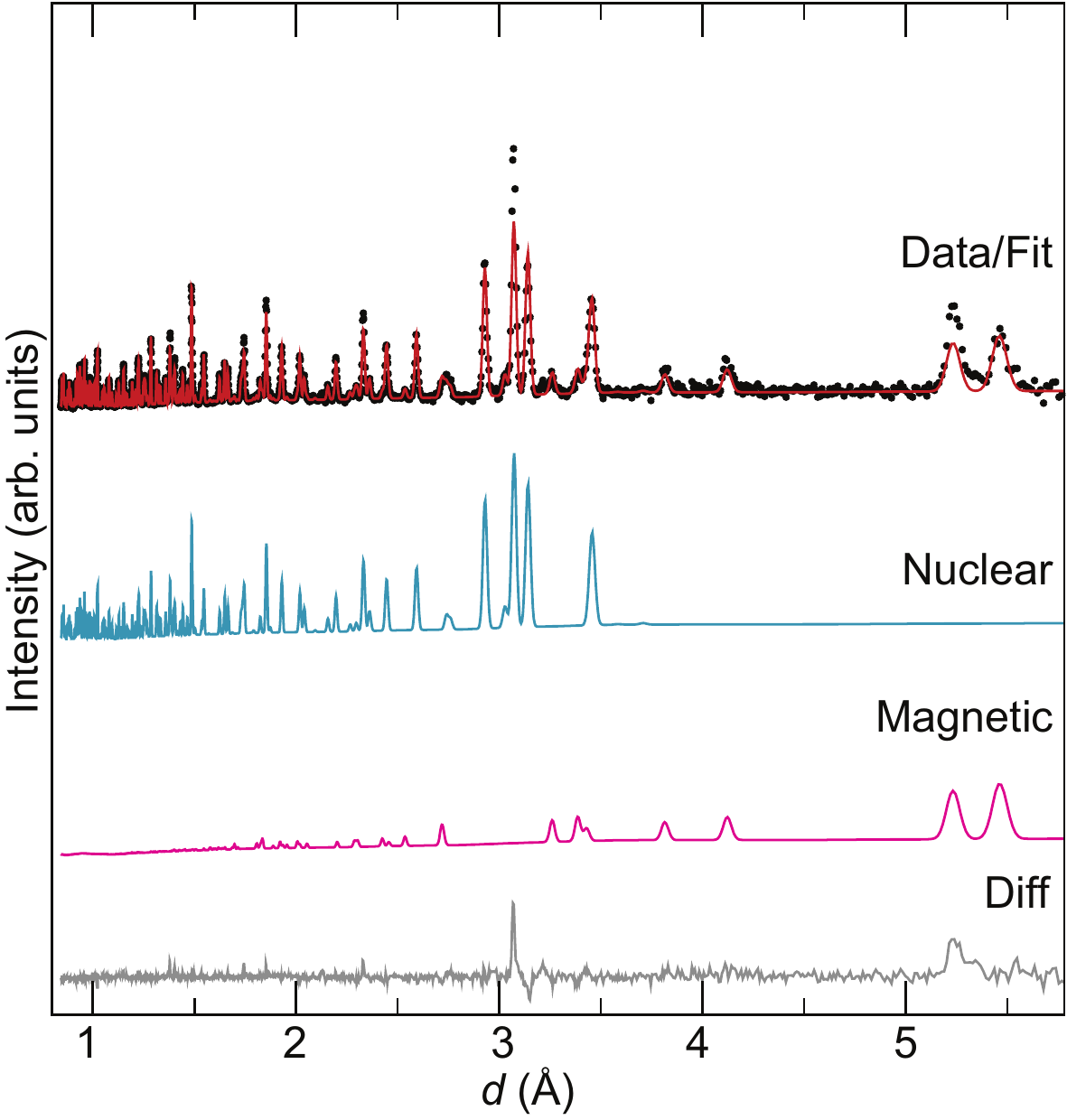}
	\caption{Rietveld refinement of neutron powder diffraction data of \kms\ collected at 10K with $k = [0.58~0~1]$.}
	\label{fig:powgenref}
\end{figure}

The spin lattice in \kms\ and \kmse\ is comprised of isosceles triangles, with two nearest-neighbor (\Jinter$_1$) and four next-nearest-neighbor interactions (\Jinter$_2$) as shown in Figure \ref{fig:unitcell}(c). The relative strengths of interchain exchange interactions \Jinter$_1$ and \Jinter$_2$, and \Jintra\ and anisotropy terms in the Hamiltonian determine the final magnetic ordering of the system. 
Such a triangular lattice can form a simple non-frustrated square lattice if 
\Jinter$_1$ = 0 and \Jinter$_2 \neq 0$, a strongly frustrated triangular lattice if \Jinter$_1~\approx$~\Jinter$_2~\neq~0$, and a decoupled spin chain if \Jinter$_1$ $\neq$ 0 and \Jinter$_2~= 0$. Overall, the magnetic frustration in \kms\ manifests itself in the form of incommensurability along $a$ direction. 
A similar ground state to \kms\ is seen in the frustrated quasi-1D triangular lattice antiferromagnetic systems CsNiCl$_3$, CsCoCl$_3$, NaFeGe$_2$O$_6$,\cite{ding2018unraveling} 
and Li$_2$NiW$_2$O$_8$. \cite{ranjith2016commensurate}
In the case of Li$_2$NiW$_2$O$_8$, strong anisotropy can overcome the effect of
frustration and stabilize a collinear magnetic order. 
Collinear magnetic structures are also observed in monoclinic compounds, 
where the unique angle $\beta$ removes the interchain frustration,
as seen in TlFeS$_2$ and TlFeSe$_2$. \cite{seidov2001magnetic}
Comparison of neutron powder diffraction patterns for \sse\ compounds from Figure \ref{fig:npd} reveal similar magnetic ordering in all compositions and the cycloid model is proposed as the magnetic structure in these compounds.

\subsection{Similar \Tn\ in \sse }

\begin{figure}
	\centering\includegraphics[width=0.95\columnwidth]{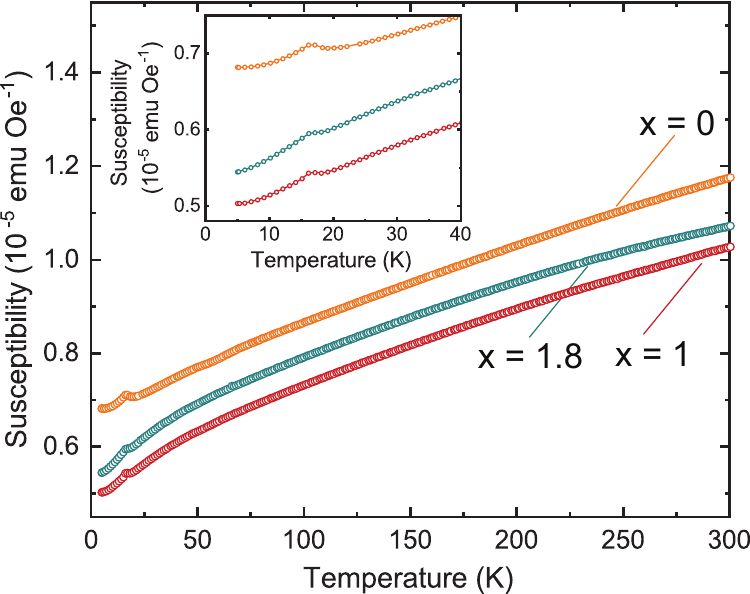}
	\caption{Magnetic susceptbility for three different compositions in K$_2$MnS$_{2-x}$Se$_x$ }
	\label{fig:mpms} 
\end{figure}

The determination of $T_N$ is more precise in these compounds when viewed by magnetic
susceptibility, single-crystal neutron diffraction, and heat capacity.
Magnetic susceptibility measurements for $x =0.0, 1.0$ and $1.8$ are displayed in Figure \ref{fig:mpms}. 
The magnetic susceptibilities of quasi-1D magnets exhibit
two characteristic maxima: a broad maximum at high 
temperatures associated with the short-range intrachain order in 1D, 
and a sharper kink at low-temperatures associated with the interchain 
transition into a 3D magnetically ordered phase. The 3D-ordering peak ($T_N$) appears at $\sim 17$~K for $x = 0$, $1$ and $1.8$.
However, our susceptibility measurements up to 300~K do not approach
the region of maximum susceptibility due to intrachain coupling,
as is typical for related systems with strong 1-D order:
KFeS$_2$ and CsFeS$_2$ display maxima at $T = 565$ and 800~K, respectively. \cite{tiwary1997regular}
Other compounds, such as TlFeS$_2$, TlFeSe$_2$, and RbFeSe$_2$, show a linear increase in susceptibility above \Tn\ and no signs of saturation, which may arise from 
delocalization of $d$ electrons due to small intrachain Fe-Fe distances. A small degree of itinerancy and hence one-dimensional metallic behavior is expected but has never been shown in these ternary iron chalcogenides.
Difficulties in handling such fragile, fiber-like crystals and the presence of defects and mechanical breaks in the crystals complicate verification of the metallic nature at a microscopic level. \cite{seidov2016magnetic, seidov2001magnetic} A linear increase in susceptibility has also been observed in two-dimensional metallic layered iron pnictides in the paramagnetic regime, which may arise from the tendency to exhibit moments that have mixed local and itinerant character.\cite{yin_kinetic_2011}
The \Mn-\Mn distance is $\sim$3.1~\AA\ in \kms\, while the shortest Mn-Mn distance in metallic manganese is 2.67~\AA, so some degree of itinerancy can be expected in \kms\ and \kmse.

\begin{figure*}
	\centering\includegraphics[width=2\columnwidth]{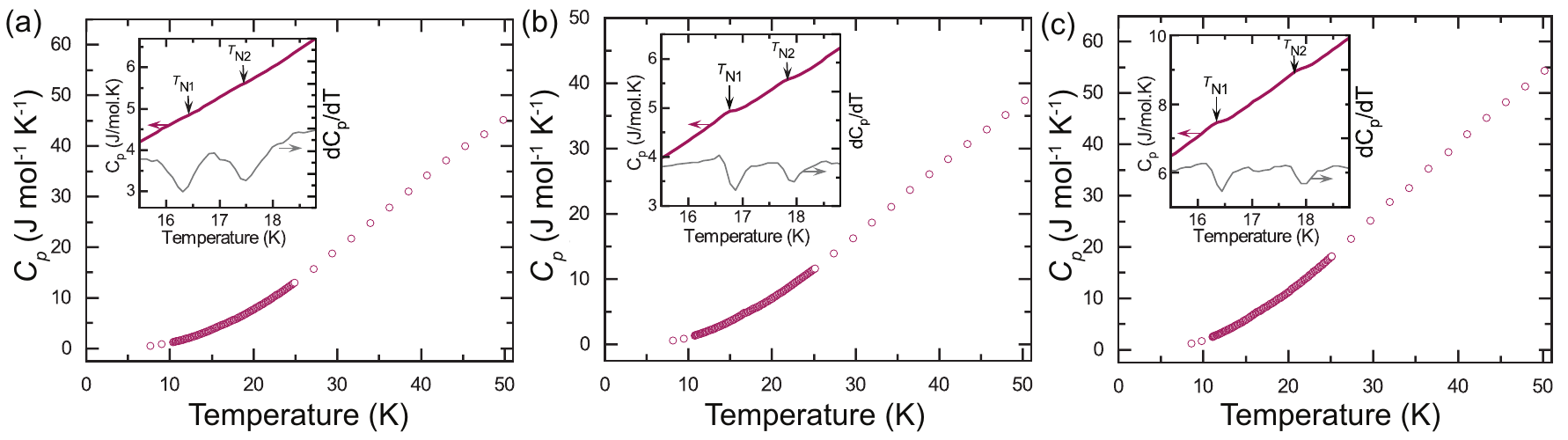}
	\caption{Heat capacity data for samples with $x=0$, 1, and 1.8 in \sse\ (a-c, respectively)
		show two closely-spaced transitions, likely corresponding to an initial interchain ordering along the shortest direction ($a$ axis), followed by full 3D ordering.
	}
	\label{fig:heatcapacity} 
\end{figure*}

In the absence of high-temperature susceptibility data showing the broad maximum and paramagnetic behavior following the Curie-Weiss law, existing models for 1D magnetic chains such as the Wagner-Friedberg model, the Rushbrooke-Wood model, the Emori model cannot be used to estimate the value of intrachain coupling as in KFeS$_2$ and CsFeS$_2$. \cite{wagner1964linear,rushbrooke1958curie,emori1969magnetic} If we assume very small degree of itinerancy in \kms\ and \kmse\ and that they reach a maximum at higher temperature, results compiled by De Jongh and Miedema can be used to put a lower bound on intrachain coupling constant $J$ using \Tmax/$J$~$\approx$~10.6, giving \Jintra$>28$~K since \Tmax~$>300$~K. \cite{de1974experiments} 

\Jintra\ is governed by the direct and super-exchange interactions between \Mn-\Mn\, while \Jinter, on the other hand, is governed by complicated super-super exchange interactions via Mn--S/Se..S/Se--Mn pathways. While the direct exchange is only governed by distance between \Mn-\Mn, the higher levels of exchange interactions depend on two opposing factors: the geometric effect of a larger bridging anion (Se) causing an increase in \Mn-\Mn\ distance, and decrease in polarizability of the bridging anion.  
The final magnetic ordering temperature \Tn\ can be conjectured based on the degree of frustration that arises from the relative strengths of \Jinter$_1$ versus \Jinter$_2$, \Jintra\, and anisotropy terms in the Hamiltonian. 
The balance between these complex terms is surprisingly robust in the case of \sse, since there is no discernible change in \Tn\ across the substitution range.

\subsection{Signature of a two-step magnetic transition in \sse}

%Heat capacity data measured for $x=0$, 1, 1.8 are shown in Figure \ref{fig:heatcapacity} from $T = 6$~K to $T = 45$~K.
%% Lattice contributions were not subtracted to estimate the magnetic contribution and Debye temperatures were also not estimated in the absence of high-temperature data. 
%A two-step transition can be observed in all three samples, which is more evident
%in the first temperature derivatives shown in the insets. 
%The transitions can be noted as two distinct N\'{e}el temperatures
%for each composition:  $T_{N1} = 16.9$~K and $T_{N2} = 18.0$~K for $x = 0$, $T_{N1} = 16.9$~K and $T_{N2} = 18.0$~K for $x = 1$, 
%and $T_{N1} = 16.8$~K and $T_{N2} = 17.9$~K for $x = 1.8$. 
%Neutron diffraction confirms that at least one of these transitions is magnetic 
%in nature, but it is most likely that they correspond to consecutive orderings
%of neighboring chains in \sse, first along the short $\langle 100 \rangle$
%interchain direction, then along the slightly longer $\langle 110 \rangle$
%direction.
%\dps{need to cite other quasi-1D systems where this happens}
%This case is more likely than two-step magnetic transitions in quasi-1D compounds
%that can arise from ordering of magnetic ions at inequivalent crystallographic sites
%(Fe$_3$PO$_4$$\cdot$8H$_2$O)\cite{forstat1965specific}, metamagnetic transitions, or association of a structural transition with a magnetic transition (BaFe$_2$As$_2$). \cite{krellner2008magnetic} \hll{cite}

Heat capacity data measured for $x=0$, 1, 1.8 are shown in Figure \ref{fig:heatcapacity} from $T = 6$~K to 45~K.
Two transitions at 16 and 18~K can be observed in all three samples,
most easily seen in the first temperature derivatives $dC_p/dT$. 
Neutron diffraction confirms that at least one of these transitions is magnetic 
in nature, but they likely correspond to two consecutive magnetic orderings. 
The frustration or competition between \Jinter$_1$ and \Jinter$_2$ with anisotropy 
is typically invoked to explain the splitting of $T_N$ into two, 
as observed in $ABX_3$-type compounds, and is relevant here.\cite{collins1997review}
In such a case, increasing the applied magnetic field should
cause the two transitions to merge, which we observed in 
\kms\ at $H=5$~T. (Figure S1)
This consecutive-ordering case is more likely than two-step magnetic transitions in quasi-1D compounds
from ordering of magnetic ions at inequivalent crystallographic sites
(all Mn sites are equivalent here),\cite{forstat1965specific}
%(Fe$_3$PO$_4$$\cdot$8H$_2$O)
or an accompanying  structural transition (as in BaFe$_2$As$_2$).\cite{krellner2008magnetic} A two-step transition could not be observed in magnetic susceptibility measurements
(Figure \ref{fig:mpms}) because the peak in susceptibility is relatively broad, with
FWHM~$> 2$~K compared to the $\sim 1.5$~K spacing between the various $T_{N1}$ and $T_{N2}$. Further measurements probing magnetic states between \Tn$_1$ and \Tn$_2$ can fully shed light on the evolution of magnetic orderings with temperature exhibited by these compounds.

%Two-step magnetic transitions - write more about the other 1D TLAFs

%The frustration or the competition between \Jinter$_1$ and \Jinter$_2$ could be a possible explanation of splitting of \Tn\ into two. A magnetic ordering between the \Mn\ ions along $a$ could occur at \Tn$_1$ followed by a final 3D ordering resulting in a final incommensurate cycloid structure at \Tn$_2$.     

\subsection{Interchain anisotropy across many compounds in the \kms\ structure type}

\begin{figure}[b]
	\centering\includegraphics[width=0.9\columnwidth]{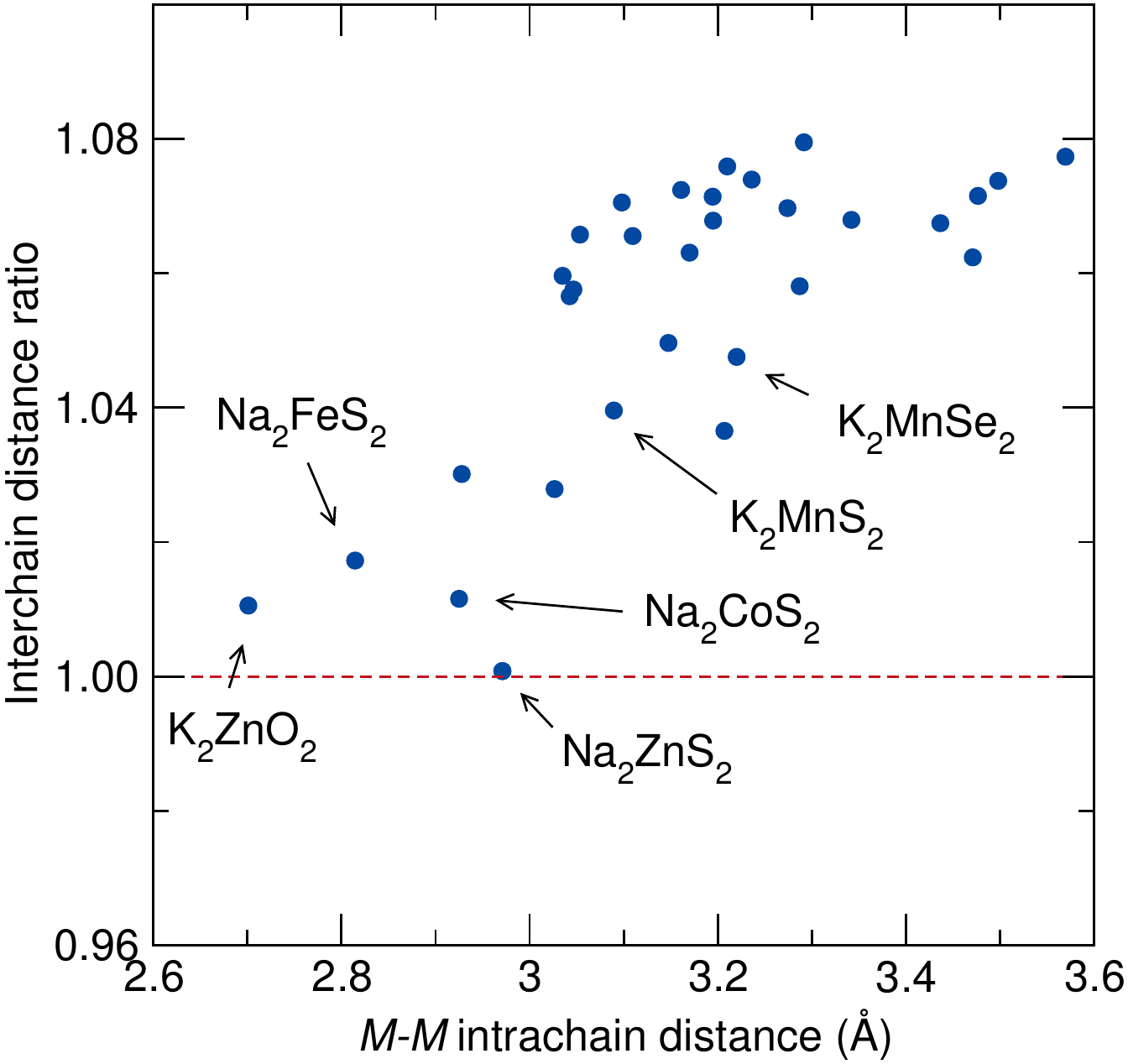}
	\caption{Ratio of the long versus short interchain distances versus the intrachain metal-metal distance for all members of the \kms\ structure type where $M$ is a cation. Compounds where the ratio = 1 have 1-D chains that lie on a nearly perfect triangular lattice.}
	\label{fig:dratio}
\end{figure} 

There are about 30 compounds of the type $A_2MX_2$ in the $Ibam$ \kms\ structure type
reported in the Inorganic Crystal Structure Database. \cite{ICSD}
Those with Mn, Co, Fe can be assumed to be magnetic, while others where $M$ = Zn, 
Si, Ge, Sn are not. 
As an arena for tuning magnetic interactions, intrachain couplings can 
produce systems that are frustrated along one direction, for example in LiCuVO$_4$, NaCu$_2$O$_2$, and CuCl$_2$, \cite{gibson2004incommensurate,capogna2005helicoidal, banks2009magnetic, capogna2010magnetic, mourigal2011ferroelectricity}
while separately, these chains interact on a triangular lattice which
we show to have much weaker ($<1000$ times) interactions. 
In the \sse\ system, the triangular lattice is anisotropic,
made of isosceles triangles where the long distance is approximately
4\% longer than the short distance. In the interest of tuning these
distances to further frustrate the system, an ideal triangular lattice
would occur when the two distances are equal. At this point we believe
that the interchain ordering could be further suppressed, below the
$\sim 17$~K temperatures seen here. In Figure \ref{fig:dratio},
we plot the ratio of the long versus short interchain distance
for all known $A_2MX_2$ compounds, along with the $M$--$M$ intrachain
distance. The only compound that has been examined by neutron diffraction
is K$_2$CoS$_2$, which Bronger and Bomba noted as an antiferromagnet
with $T_N > 9.5$~K, but no data or metrics were included to evaluate the
structure solution.\cite{bronger1989} Their proposed model included Co moments pointing
along the intrachain direction, so there are clearly many degrees
of freedom that remain unexplored in this diverse system, and the 
full list of compounds with distances in ratios is given in Table S1.

%
%\subsection{Photoluminescence in \sse}
%
%\begin{figure*}
%	\centering\includegraphics[width=1.9\columnwidth]{fig-UVVis-01}
%	\caption{\label{uvvis} (a) UV-Vis (b) PL from 488 nm laser at MRL (c) PL from 266 nm laser at MRL. 1.8eV = 689nm (red); 3.05 eV = 407nm (violet); 4.25 eV = 292nm }
%\end{figure*} 

%\subsection{PL and UV-Vis}
%\begin{figure}
%	\centering\includegraphics[width=0.95\columnwidth]{fig-PL + UV-Vis}
%	\caption{\label{struc} PL }
%\end{figure} 

\section{Summary and outlook}\label{summary}

In summary, we have presented the first detailed investigation of \kms\ and \kmse\ compounds from their eponymous structure type and their magnetic solid solution \sse. We observe an incommensurate cycloid magnetic structure
% with a magnetic propagation vector $k = [0.58~0~1]$ %
for all samples in \sse, identified by single crystal neutron diffraction of \kms\ at 4~K and powder neutron diffraction of all samples at 10 and 50~K. The quasi-1D compound is best represented as a 2D triangular antiferromagnet, which results in geometric frustration of chains resulting in incommensurability along $a$, a two-step magnetic transition characteristic of differing \Jinter$_1$ and \Jinter$_2$, and the prospect for tuning these interactions via a wide array of substitution in isostructural compounds.

\section*{Acknowledgments}

We acknowledge support from the Center for Emergent Superconductivity, an Energy Frontier Research Center funded by the U.S. Department of Energy, Office of Science, Office of Basic Energy Sciences under Award Number DEAC0298CH1088. Characterization was performed in the Materials Research Laboratory Central Research Facilities, University of Illinois.
% Photoluminescence measurements were carried out in the Beckman Institute of Advanced Science and Technology, University of Illinois. 
Single crystal neutron diffraction and powder neutron diffraction were conducted at ORNL's High Flux Isotope Reactor and Spallation Neutron Source, sponsored by the Scientific User Facilities Division, Office of Basic Energy Sciences, U.S. Department of Energy.

\bibliography{k2mns2}

\end{document}